\documentclass[11pt,a4paper]{article}
\usepackage{jheppub}
\usepackage{amsmath}
\usepackage{multicol,bbm}
\usepackage{enumerate}
\usepackage{bibentry}

\graphicspath{{./Figs/}}

\def\beq{\begin{equation}}
\def\eeq{\end{equation}}

\def\be{\begin{equation}}
\def\ee{\end{equation}}
\def\ba{\begin{eqnarray}}
\def\ea{\end{eqnarray}}
\def\bea{\begin{eqnarray}}
\def\eea{\end{eqnarray}}
\def\eq{\begin{equation}}
\def\eqe{\end{equation}}
\def\eqa{\begin{eqnarray}}
\def\eqae{\end{eqnarray}}

\def\beqa{\begin{eqnarray}}
\def\eeqa{\end{eqnarray}}

\newcommand{\beqas}{\begin{eqnarray*}}
\def\beqa{\begin{eqnarray}}
\def\eeqa{\end{eqnarray}}
\def\beq{\begin{equation}}
\def\eeq{\end{equation}}

\newcommand{\eeqas}{\end{eqnarray*}}

\title{Double Soft Theorems in Gauge and String Theories}
\author[a]{Anastasia Volovich,}
\author[b]{Congkao Wen,}
\author[a]{Michael Zlotnikov}
\affiliation[a]{Brown University
Department of Physics
182 Hope St, Providence, RI, 02912} 
\affiliation[b]{I.N.F.N. Sezione di Roma ``Tor Vergata",
Via della Ricerca Scientifica, 00133 Roma, Italy}
\emailAdd{anastasia\_volovich@brown.edu, michael\_zlotnikov@brown.edu, Congkao.Wen@roma2.infn.it} 

\abstract{We investigate the tree-level S-matrix in gauge theories and open superstring theory with several soft particles. 
We show that scattering amplitudes with two or three soft gluons of non-identical helicities behave universally in the limit, with multi-soft factors
which are not the product of individual soft gluon factors.
The results are obtained from the BCFW recursion relations in four dimensions, and further extended to arbitrary dimensions using the CHY formula. 
We also find new soft theorems for double soft limits of scalars and fermions in ${\cal N}=4$  and pure ${\cal N}=2$ SYM.
 Finally, we show that the double-soft-scalar theorems can be extended to open superstring theory without receiving any $\alpha'$ corrections. }

\begin{document}
\maketitle


\section{Introduction}

Recently there has been a resurrection of interest in studying various low energy limits of scattering amplitudes.
Of particular interest are situations which exhibit universal behavior; that is, when the limiting behavior of an
amplitude factors into a product of a universal ``soft factor'' times a lower-point amplitude independent of the
soft particles.  Such cases are  called ``soft theorems'', the most famous
of which may be Weinberg's classic soft (photon, gluon, or graviton) theorems \cite{Weinberg}.  
Other theorems include  ~\cite{LBK, GJ, Bell:1969yw, DelDuca:1990gz} as well as, 
much more recently, the subleading and sub-subleading graviton theorems of Cachazo and Strominger
~\cite{CS} (see \cite{other} for further developments and applications).

Strominger and collaborators \cite{BMS-strominger} have argued that all of the known soft and subleading soft theorems
may be understood as consequences of large gauge transformations. That is, transformations which fall off sufficiently rapidly
at infinity such that they must be considered consistent with the asymptotic boundary conditions defining the theory, while sufficiently
slowly that they act nontrivially on asymptotic scattering states.
In the case of gravity, the relevant ``gauge transformations'' are of course diffeomorphisms, and the relevant
asymptotic symmetry group (in four-dimensional Minkoswki space)
is the Bondi, van der Burg,  Metzner, Sachs group \cite{BMS, BT}.  It has been shown using the CHY scattering
equations \cite{CHY} that the subleading and sub-subleading graviton soft theorems hold for tree-level
graviton amplitudes in any number of space-time dimensions, suggesting that an analog of the BMS symmetry should be relevant
more generally \cite{Ddim,BMS-strominger}. Perturbative theories at null infinity realizing these symmetries have been proposed in \cite{null}.
 The issues regarding possible loop corrections to the subleading soft theorems were studied in~\cite{loops}.

Double-soft limits (where two particles are taken to have very low energy)
 have also received a lot of attention in the literature, both in the earlier works \cite{JS, weinberg2} and more recently.
For example, Arkani-Hamed et.~al.~\cite{nima} have shown that the double soft limit of scalars in ${\cal N}=8$ supergravity exhibit
the expected $E_{7(7)}$ symmetry of the scalar moduli space, in a manner analogous to the classic soft-pion theorem
of \cite{adler, weinberg2}. This result was recently extended to the four-dimensional supergravity theories with ${\cal N} < 8$ supersymmetry, and the $\mathcal{N}=16$ supergravity in three dimensions in~\cite{congkao2}. Furthermore, supergravity amplitudes in both four and three dimensions with two soft fermions were studied in~\cite{congkao1},
and new soft theorems were proposed. 
New double-soft leading and subleading theorems for scalars (and leading for photons) were also studied in various theories such as DBI, 
Einstein-Maxwell-scalar, NLSM, and Yang-Mills-scalar in~\cite{CHY-double}. Kac-Moody
structure has been found for the four dimensional Yang-Mills at null infinity~\cite{He:2015zea}, where double soft limits play another important role.

In this paper we derive several new soft theorems for tree-level scattering amplitudes in gauge and string theories with more than one soft particle. We derive the universal behavior of amplitudes with two or three soft gluons. It is known that when the soft gluons have identical helicities, the result can be obtained simply by setting the gluons to be soft one by one, thus we focus on the non-trivial cases when the soft gluons have different helicities. Indeed we find that for these cases the soft factors are a product of the individual soft-gluon factors with certain non-trivial corrections. We first derive theorems from the BCFW formula in four dimensions \cite{BCFW}, and
further extend our results with double-soft gluons for gauge theories in any number of dimensions by using CHY formula \cite{CHY}. We check that our results are consistent with the fact that if the soft limit is taken in order then the soft factors reduce to a product of the single-soft factors given by Weinberg. We also note that, in contrast to the gluon case, amplitudes with multiple soft gravitons can always be obtained by simply taking the gravitons to be soft one by one. 

We then proceed to  study amplitudes in ${\cal N}  = 4$ and pure ${\cal N} = 2$ Super Yang-Mills theory (SYM) with two soft scalars or two soft fermions. We find 
that the double soft behavior is governed by R-symmetry generators acting on a lower-point amplitude, 
resembling the results of supergravity theories found in~\cite{nima,congkao2,congkao1}, although the vacuum structure of SYM is quite different from that of supergravity theories. Finally, we consider double-soft scalars in the open superstring theory. Unlike the double-soft-scalar theorem in $\mathcal{N}=8$ supergravity, which would receive $\alpha'$ corrections if one tried to extend it to closed superstring theory, we find that open superstring amplitudes satisfy exactly the same double-soft-scalar theorem of SYM at $\alpha'=0$. Given the similarity of the double-soft theorems of SYM and those of supergravity theories, it would be very interesting to understand if any of these theorems could have an interpretation as hidden symmetries. 

The paper is organized as follows: In section \ref{section:doublegluons} we derive the double-soft-gluon theorem for tree-level amplitudes in gauge theories using the BCFW recursion relations formula
(\ref{eq:tnjw}), which may be recast into a different form (\ref{eq:tnjw2}), and we further extend the results to arbitrary dimensions resulting in formula (\ref{CHY2gluon}). Then, in section \ref{section:triplegluons} amplitudes with three soft gluons are considered. 
In the following section \ref{section:gravity} we comment on multi-soft gravitons.
Subsequently, in section \ref{section:SYM}, we explore the universal behavior of amplitudes with two soft scalars or two soft fermions in supersymmetric gauge theories, including $\mathcal{N}= 4$ SYM as well as pure $\mathcal{N}=2$ SYM, with main results given by (\ref{doublescalars}) and (\ref{doublefermions}). Finally, in section~\ref{section:string}, we prove that the newly discovered double-soft-scalar theorem in SYM can be extended to the open superstring theory without any $\alpha'$ corrections. 

\bigskip

{Note added}: After finishing this work, we became aware of a related work by Klose, McLoughlin, Nandan, Plefka and  Travaglini,
which has some overlap with our paper \cite{klose}.

\section{Double-soft gluons} \label{section:doublegluons}

\subsection{Double-soft gluons from BCFW recursions}

We start by considering color-stripped amplitudes in gauge theories with two adjacent gluons taken to be soft. It is straightforward to see that if the two gluons have the same helicity, then the two gluons may be taken soft one at a time. Moreover it is evident from
\beqa \label{2ppgluons}
\lim_{p_2 \to 0} \lim_{p_1 \to 0} A(1^+, 2^+, 3, \ldots, n) 
&\rightarrow& 
{ \langle n2\rangle \over \langle n1\rangle \langle 12\rangle  }
{ \langle n3\rangle \over \langle n2\rangle \langle 23\rangle  } A(3, \ldots, n) \cr
\lim_{p_1\to 0} \lim_{p_2 \to 0} A(1^+, 2^+, 3, \ldots, n) 
&\rightarrow&
{ \langle 13\rangle \over \langle 12\rangle \langle 23\rangle  }
{ \langle n3\rangle \over \langle n1\rangle \langle 13\rangle  } A(3, \ldots, n)
\eeqa
that the result is independent of the order in which the two gluons are taken soft. 
 
A similar simple calculation shows that if the two gluons have different helicities, then the result cannot be given by a product of two single soft factors obtained by taking the gluons to be soft one by one. Therefore this is the non-trivial case we are interested in, namely we would like to study the amplitude $A(1^+, 2^-, 3, \ldots, n)$ in the double-soft limit 
\beq
p_{1,2} \rightarrow \tau p_{1,2} {\rm  ~~~~~with~~~~~} \tau \rightarrow 0. 
\eeq
We will use the standard spinor-helicity formalism for the four-dimensional massless particles throughout this paper:
\beq
 p_{\alpha, \dot{ \alpha} } = \lambda_{\alpha} \tilde{\lambda}_{\dot{ \alpha}} ,~~~
  \langle i j \rangle = \epsilon_{\alpha \beta  } \lambda^{\alpha}_i \lambda^{\beta}_j,  ~~~
  [ij] = \epsilon_{\dot{ \alpha} \dot{ \beta}  } \tilde{\lambda}^{\dot{ \alpha}}_i \tilde{\lambda}^{\dot{ \beta}}_j
\eeq
and realize the soft limit by taking 
\beq
 \lambda_{1,2} \rightarrow \sqrt{\tau} \lambda_{1,2}  {\rm~~~~and~~~~} \tilde{\lambda}_{1,2} \rightarrow \sqrt{\tau} \tilde{\lambda}_{1,2}.
 \eeq
Using the BCFW recursion relations \cite{BCFW}, it is straightforward to see that the two dominant diagrams that contribute in the limit at hand are
\beq
\vcenter{\hbox{\includegraphics[scale=0.6]{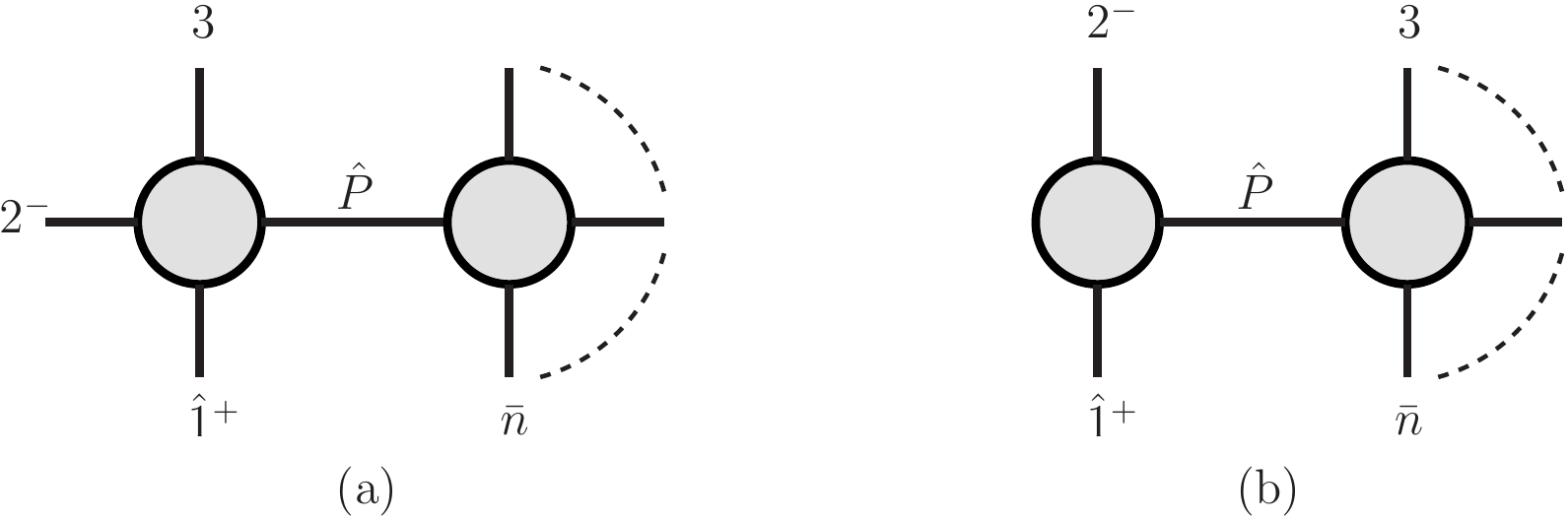}}}
\eeq
with the following BCFW shifts
\beq
\lambda_{\hat{1}} = \lambda_1 + z \lambda_n \, ,  
\quad 
\tilde{\lambda}_{\bar{n}} = \tilde{\lambda}_n - z \tilde{\lambda}_1 \, .
\eeq
Let us now analyse the two contributions separately. First for diagram (a) we have
\beq
A_{\rm (a)} = {[13]^4 \over [12][23][3 \hat{P}][\hat{P} 1] s_{123} } A_{n-2} 
  \rightarrow  {[13]^3 \langle n 3\rangle \over [12][23] \langle n|1+2 |3]  s_{123} } A_{n-2} \, ,
\eeq
where we have used the fact that $\hat{P} \to  p_3$ in the limit; hence this result is independent of whether particle $3$ has positive or negative helicity (in the above calculation we have chosen it to be positive). Now, the second diagram (b) gives 
\beqa
A_{\rm (b)} = { [\hat{P} 1]^3 \over [12][2 \hat{P}] s_{12} } A_{n-1}( \hat{P}, 3, \ldots, \bar{n} ) \rightarrow 
{ [\hat{P} 1]^3 \over [12][2 \hat{P}] s_{12} }  { [n3] \over [ n \hat{P} ][ \hat{P} 3]  } A_{n-2}(3, \ldots, n ) \, ,
\eeqa
where in the second expression we have used the fact that $p_{\bar{n}} = p_n$ in the limit, and we also applied the single-soft theorem for the soft leg $\hat{P}$. After some simplification, we find
\beqa
A_{\rm (b)} \rightarrow  
{ \langle n2\rangle^3 [n3] \over  \langle n1\rangle \langle 12\rangle \langle n| 1+2 |3] s_{n12}} A_{n-2} \, .
\eeqa
Adding the contributions from the two diagrams together, we obtain the final result for two soft gluons having different helicities,
\beqa 
\label{eq:tnjw}
\lim_{p_1 \sim p_2 \to 0} A(1^+, 2^-, 3, \ldots, n) \rightarrow 
{1 \over \langle n| 1+2 |3] }
\left( {[13]^3 \langle n3\rangle \over [12][23] s_{123}}
+ { \langle n2\rangle^3 [n3] \over \langle n1\rangle \langle 12\rangle s_{n12}} \right) A_{n-2} \,.
\eeqa
As typical for amplitudes computed from the BCFW recursion relations, the result contains a spurious pole ${1 \over \langle n| 1+2 |3] }$. We will show that it indeed cancels out between the two terms at leading order of the soft limit, as should be the case. Now, if on the other hand we take the soft limit in succession, namely say take $p_1$ to be soft first, then the first term in the soft factor is subleading, and the second term simplifies to
\beq
{1 \over \langle n| 1+2 |3] }
\left( {[13]^3 \langle n3\rangle \over [12][23] s_{123}}
+ { \langle n2\rangle^3 [n3] \over \langle n1\rangle \langle 12\rangle s_{n12}} \right) 
\rightarrow 0+ { \langle n2\rangle \over \langle n1\rangle \langle 12\rangle } { [n3] \over [32][2n]} \, ,
\eeq
which is precisely the product of two soft factors of a positive gluon and a negative gluon, with the positive gluon $p_1$ being taken soft first. 

Although the above result~(\ref{eq:tnjw}) is very compact and nicely reduces to a product of two soft factors, if we take the soft limits in succession, it is specific to four dimensions and as a natural property of using the BCFW recursion it contains a spurious pole. In the next section we will use the CHY formula for pure Yang-Mills tree level scattering amplitudes \cite{CHY} to derive a further formula for the universal double-soft-gluon factor. This result will be valid in any dimension, for any helicity combination of the soft gluons, and it will be manifestly free of unphysical poles. When the two soft gluons have opposite helicity, the comparison of the result obtained from the CHY formula and (\ref{eq:tnjw}) will yield agreement and provide us with the intuition to recast the above into the following equivalent form
\begin{align} \label{eq:tnjw2}
\lim_{p_1 \sim p_2 \to 0} A(1^+, 2^-, 3, \ldots, n) \rightarrow \frac{\langle n2\rangle}{\langle n1\rangle\langle 12\rangle}\frac{[13]}{[12][23]}\left(1
+\frac{ \langle n1\rangle [13] \langle 32\rangle}{s_{123}\langle n2\rangle}
+\frac{[1n]\langle n2\rangle[23]}{s_{n12}[13]}\right)A_{n-2}\, .
\end{align}
Therefore, we see that the alternating helicity double soft gluon factor is composed of the product of two single soft gluon factors plus a non-trivial correction.

\subsection{Double-soft gluons from CHY}
As we mentioned earlier, in this section we will reconsider the double-soft-gluon limit making use of the CHY formula for tree-level scattering amplitudes in pure Yang-Mills, valid in arbitrary dimensions~\cite{CHY}. The CHY formula for an $n$-point gluon scattering amplitude is given by
\begin{align}
\mathcal{A}_{n}=\int\left(\prod_{{c=1}\atop{c\neq p,q,r}}^{n}d\sigma_c\right)\frac{(\sigma_{pq}\sigma_{qr}\sigma_{rp})(\sigma_{ij}\sigma_{jk}\sigma_{ki})}{\sigma_{12}\sigma_{23}...\sigma_{n,1}}\left(\prod_{{a=1}\atop{a\neq i,j,k}}^{n}{\delta\left(f_a\right)}\right)\frac{2(-1)^{m+w}}{\sigma_{mw}} {\rm Pf}\left(\Psi^{m,w}_{m,w}\right),\notag
\end{align}
where $\sigma_{ij}\equiv(\sigma_i-\sigma_j)$ and $f_a=\sum_{b\neq a}\frac{k_a\cdot k_b}{\sigma_{ab}}$. Upper and lower indices on the matrix $\Psi$ denote removed columns and rows respectively. The indices $p,q,r,i,j,k,m$ and $w$ can be fixed arbitrarily without changing the result. The $2n\times2n$ dimensional matrix $\Psi$ is given by
\begin{align}
\Psi=\left({{A}\atop{C}}~{{-C^T}\atop{B}}\right),\notag
\end{align}
where the $n\times n$ dimensional sub-matrices are
\begin{align}
A_{ab}=\left\{{{\frac{k_a\cdot k_b}{\sigma_{ab}}}\atop{0}}{{,~a\neq b}\atop{,~a=b}}\right.~~~,~~~B_{ab}=\left\{{{\frac{\epsilon_a\cdot \epsilon_b}{\sigma_{ab}}}\atop{0}}{{,~a\neq b}\atop{,~a=b}}\right.~~~,~~~C_{ab}=\left\{{{\frac{\epsilon_a\cdot k_b}{\sigma_{ab}}}\atop{-\sum_{{c=1}\atop{c\neq a}}^{n}\frac{\epsilon_a\cdot k_c}{\sigma_{ac}}}}{{,~a\neq b}\atop{,~a=b}}\right..\notag
\end{align}
Here $k_a^\mu$ are external leg momenta, and the $\epsilon_a^\mu$ are corresponding polarization vectors. The product of delta functions enforces the scattering equations and saturates all integrals. With this the integration reduces to a sum over all solutions to the scattering equations.\\
\indent We want to make the external gluon momenta $k_{1}^\mu$ and $k_{2}^\nu$ soft by substituting $k_{1}^\mu\rightarrow \tau k_{1}^\mu$ and $k_{2}^\nu\rightarrow \tau k_{2}^\nu$ and considering $\tau\rightarrow 0$. It is essential to send both momenta to zero simultaneously in order to capture the double soft factor structure. We choose not to erase indices $(1)$ and $(2)$. With this we have to isolate the extra terms in $\mathcal{A}_{n}$ as compared to $\mathcal{A}_{n-2}$ and integrate out $\sigma_{1}$ and $\sigma_{2}$. While doing so we will only keep the leading contribution in the $\tau\rightarrow 0$ limit, to obtain the leading double soft gluon factor.

First we notice that at leading order in $\tau$ the entire $\sigma_{1}$ and $\sigma_{2}$ dependence in $\mathcal{A}_{n}$ apart from the pfaffian ${\rm Pf}\left(\Psi^{m,w}_{m,w}\right)$ is contained in
\beq
\label{isolatesn1sn2}
\int d\sigma_{1}d\sigma_{2}\frac{\sigma_{n,3}}{\sigma_{n,1}\sigma_{1,2}\sigma_{2,3}}\delta(f_{1})\delta(f_{2}).
\eeq
Another $\sigma_{n,3}$ term in the denominator is suppressed, which will help restore the proper Parke-Taylor factor for the $(n-2)$-point amplitude case. As in \cite{CHY-double}, we can make the convenient variable transformation
\begin{align}
&\sigma_{1}=\rho-\xi/2,~~~\sigma_{2}=\rho+\xi/2,\notag\\
&d\sigma_{1}d\sigma_{2}\delta(f_{1})\delta(f_{2})=-2d\rho d\xi\delta(f_{1}+f_{2})\delta(f_{1}-f_{2}),
\end{align}
and immediately integrate out $\delta(f_{1}-f_{2})$ using the variable $\xi$. This will introduce a summation over all solutions $\xi$ for the equation $f_{1}-f_{2}=0$, and an overall factor of $1/F(\xi)$, where
\begin{align}
&F(\xi)=\frac{d}{d\xi}(f_{1}-f_{2})\\
&=\frac{1}{2}\frac{1}{k_{1}\cdot k_{2}}\left(\sum_{b=3}^n\left(\frac{k_{1}\cdot k_{b}}{\rho-\frac{\xi}{2}-\sigma_b}-\frac{k_{2}\cdot k_{b}}{\rho+\frac{\xi}{2}-\sigma_b}\right)\right)^2+\frac{1}{2}\sum_{c=3}^n\left(\frac{\tau k_{1}\cdot k_{c}}{\left(\rho-\frac{\xi}{2}-\sigma_c\right)^2}+\frac{\tau k_{2}\cdot k_{c}}{\left(\rho+\frac{\xi}{2}-\sigma_c\right)^2}\right).\notag
\end{align}
 Here we used that on the support of $f_{1}-f_{2}=0$ we can always substitute
\begin{align}
\label{xiexact}
\xi=\tau\frac{2k_{1}\cdot k_{2}}{\sum_{b=3}^n\left(\frac{k_{1}\cdot k_{b}}{\rho-\frac{\xi}{2}-\sigma_b}-\frac{k_{2}\cdot k_{b}}{\rho+\frac{\xi}{2}-\sigma_b}\right)}.
\end{align}
Making use of this, (\ref{isolatesn1sn2}) becomes
\begin{align}
\label{beforecontour}
\sum_{{\rm solutions}\,\xi} \int d\rho\frac{\delta(f_{1}+f_{2})}{\tau (k_{1}\cdot k_{2})F(\xi)}\sum_{b=3}^n\left(\frac{k_{1}\cdot k_{b}}{\rho-\frac{\xi}{2}-\sigma_b}-\frac{k_{2}\cdot k_{b}}{\rho+\frac{\xi}{2}-\sigma_b}\right)\frac{\sigma_{n,3}}{\sigma_{n,1}\sigma_{2,3}}.
\end{align}
Before we rewrite $\int d\rho \delta(f_{1}+f_{2})$ as a contour integral over poles and deform the contour as usual, we should also extract the extra terms depending on $\rho$ and $\xi$ from the pfaffian factor ${\rm Pf}\left(\Psi^{m,w}_{m,w}\right)$ in order to reduce it to the $(n-2)$-point amplitude case. To do that, we will use the recursive definition of a pfaffian for an anti-symmetric $2n\times 2n$ matrix $A$:
\begin{align}
{\rm Pf}\left(A\right)=\sum_{{j=1}\atop{j\neq i}}^{2n}(-1)^{i+j+1+\theta(i-j)}a_{ij}{\rm Pf}\left(A^{ij}_{ij}\right),
\end{align}
where $a_{ij}$ is an element of matrix $A$, $\theta(x)$ is the Heaviside step function, and index $i$ can be chosen arbitrarily. If rows and/or columns are missing from matrix $A$ before the expansion is applied, the respective indices have to be skipped in the summation. Since we are ultimately interested in the leading double soft gluon factor, for convenience we will only keep the leading in $\tau$ terms in the expansion of ${\rm Pf}\left(\Psi^{m,w}_{m,w}\right)$. In order to isolate the leading terms, we recall that the summation over the solutions $\xi$ in (\ref{beforecontour}) features two types of solutions: non-degenerate solutions for which $\xi=O(1)$, and a unique degenerate solution for which $\xi=O(\tau)$ \cite{CHY-double}.\\
\indent Let us first consider the non-degenerate (nd) solutions. The (nd) case is non-trivial, since equation (\ref{xiexact}) seemingly has to be solved in $\xi$ for the full non-linear constraint imposed by the scattering equations involved, yet polynomial roots can be obtained in closed form for low degree polynomials only. It is possible to derive non-degenerate solution contributions in this case employing a somewhat cumbersome procedure. However, this will not be required in the following and will be addressed in more generality in a future work. Instead, investigation of the soft factor integrand reveals that the necessity of non-degenerate solutions computation can be avoided here at the expense of fixing a particular polarization gauge for the two gluons going soft.\footnote{The lost gauge invariance in the final result is recovered once we convert it to spinor helicity formalism.} This argument works as follows.\\
In the (nd) case it is straightforward to see that the only leading term in the pfaffian expansion is given by
\begin{align}
\label{ndpfaffian}
{\rm Pf}\left(\Psi^{m,w}_{m,w}\right)_{({\rm nd})}&=-C_{1,1}C_{2,2}{\rm Pf}\left(\Psi^{m,w,1,2,n+1,n+2}_{m,w,1,2,n+1,n+2}\right)+O(\tau)\notag\\
&=-\sum_{b=3}^n\frac{\epsilon_{1}\cdot k_b}{\rho-\frac{\xi}{2}-\sigma_b}\sum_{c=3}^n\frac{\epsilon_{2}\cdot k_c}{\rho+\frac{\xi}{2}-\sigma_c}{\rm Pf}\left(\Psi'^{m,w}_{m,w}\right)+O(\tau),
\end{align}
where for convenience we define the abbreviation
\begin{align}
\label{psiabbr}
{\rm Pf}\left(\Psi'^{m,w}_{m,w}\right)\equiv{\rm Pf}\left(\Psi^{m,w,1,2,n+1,n+2}_{m,w,1,2,n+1,n+2}\right).
\end{align}
Combining (\ref{beforecontour}) with (\ref{ndpfaffian}), writing $\int d\rho \delta(f_{1}+f_{2})$ as a contour integral
\begin{align}
\label{deltacontour}
\int d\rho \delta(f_{1}+f_{2})\rightarrow \oint \frac{d\rho}{2\pi i} \frac{1}{f_{1}+f_{2}},
\end{align}
and deforming the contour to wrap around all other poles in $\rho$ instead, immediately reveals that there is no pole at infinity and the only residues come from poles at $(\rho+\xi/2-\sigma_3)\rightarrow 0$ and/or $(\rho-\xi/2-\sigma_n)\rightarrow 0$ due to the term ${\sigma_{n,3}}/({\sigma_{n,1}\sigma_{2,3}})$ remaining from the Parke-Taylor factor. Keeping (\ref{ndpfaffian}) in mind, this tells us that for any of the non-degenerate solutions $\xi_{(nd)}$, at leading order in $\tau$ these residues will always be proportional to $\epsilon_{2}\cdot k_3$ and/or $\epsilon_{1}\cdot k_n$. Therefore, we select the following polarization gauge for the external legs going soft
\begin{align}
\label{epsilongauge}
\epsilon_{2}\cdot k_3=0~~~~~,~~~~~\epsilon_{1}\cdot k_n=0.
\end{align}
In this gauge all the non-degenerate solution contributions to the leading double soft gluon factor vanish, such that we can concentrate on the degenerate solution only.

Now we compute the degenerate (d) solution contribution. Using (\ref{xiexact}) we can straightforwardly expand the degenerate solution $\xi_{(d)}$ to leading order
\begin{align}
\label{xileading}
\xi_{(d)}=\tau\frac{2k_{1}\cdot k_{2}}{\sum_{b=3}^n\frac{(k_{1}-k_{2})\cdot k_{b}}{\rho-\sigma_b}}+O(\tau^2).
\end{align}
All the terms appearing in (\ref{beforecontour}) are expanded to leading order in $\tau$ analogously. The expansion of the pfaffian features three leading terms in this case:
\begin{align}
\label{dpfaffian}
{\rm Pf}\left(\Psi^{m,w}_{m,w}\right)_{({\rm d})}&=\left(B_{1,2}A_{1,2}+C_{1,2}C_{2,1}-C_{1,1}C_{2,2}\right){\rm Pf}\left(\Psi'^{m,w}_{m,w}\right)+O(\tau)\notag\\
&=\frac{1}{4}\left[\left(\frac{\epsilon_{1}\cdot \epsilon_{2}}{k_{1}\cdot k_{2}}-\frac{(\epsilon_{2}\cdot k_{1})(\epsilon_{1}\cdot k_{2})}{(k_{1}\cdot k_{2})^2}\right)S^2+\right.\\
&~~~\left.+\left(\frac{\epsilon_{1}\cdot k_{2}}{k_{1}\cdot k_{2}}S-2\sum_{i=3}^n\frac{\epsilon_{1}\cdot k_i}{\rho-\sigma_i}\right)\left(\frac{\epsilon_{2}\cdot k_{1}}{k_{1}\cdot k_{2}}S+2\sum_{j=3}^n\frac{\epsilon_{2}\cdot k_j}{\rho-\sigma_j}\right)\right]{\rm Pf}\left(\Psi'^{m,w}_{m,w}\right)+\notag\\
&~~~+O(\tau),\notag
\end{align}
where $S=\sum_{b=3}^n\frac{(k_{1}-k_{2})\cdot k_b}{\rho-\sigma_b}$, and we used the abbreviation (\ref{psiabbr}). Again, we combine (\ref{beforecontour}) with (\ref{dpfaffian}), write $\int d\rho \delta(f_{1}+f_{2})$ as a contour integral (\ref{deltacontour}) and deform the contour to wrap around all other poles in $\rho$ instead. Analogously to the non-degenerate case we see that there is no pole at infinity, and the only two contributing residues come from poles at $\rho-\sigma_3\rightarrow 0$ and $\rho-\sigma_n\rightarrow 0$. Dropping ${\rm Pf}\left(\Psi'^{m,w}_{m,w}\right)$, which is part of the $(n-2)$-point amplitude and not the double soft gluon factor, both these residues are of the following type at leading order in $\tau$:
\begin{align}
R^q_{i,i+1}=&\frac{1}{2}\frac{(k_{i}-k_{i+1})\cdot k_q}{(k_{i}+k_{i+1})\cdot k_q}\left[\frac{\epsilon_{i}\cdot\epsilon_{i+1}}{k_{i}\cdot k_{i+1}}-\frac{(\epsilon_{i+1}\cdot k_{i})(\epsilon_{i}\cdot k_{i+1})}{(k_{i}\cdot k_{i+1})^2}+\right.\\
&~~~~~~~~~~~~~~~~~~~~~~~~\left.+\left(\frac{\epsilon_{i}\cdot k_{i+1}}{k_{i}\cdot k_{i+1}}-\frac{2\epsilon_{i}\cdot k_{q}}{(k_{i}-k_{i+1})\cdot k_{q}}\right)\left(\frac{\epsilon_{i+1}\cdot k_{i}}{k_{i}\cdot k_{i+1}}+\frac{2\epsilon_{i+1}\cdot k_{q}}{(k_{i}-k_{i+1})\cdot k_{q}}\right)\right].\notag
\end{align}
With this we conclude that the leading double soft gluon factor for legs $i$ and $i+1$ going soft is given by
\begin{align}
\label{iCHY2gluon}
S^{(0)}_{i,i+1}=&(R^{i+2}_{i,i+1}-R^{i-1}_{i,i+1})\\
=&\frac{1}{(k_{i}+k_{i+1})\cdot k_{i+2}}\left(\frac{1}{2}\frac{\epsilon_{i}\cdot\epsilon_{i+1}}{k_{i}\cdot k_{i+1}}(k_{i}-k_{i+1})\cdot k_{i+2}-\frac{\epsilon_{i+1}\cdot k_{i}}{k_{i}\cdot k_{i+1}}\epsilon_{i}\cdot k_{i+2}\right)\notag\\
&-\frac{1}{(k_{i}+k_{i+1})\cdot k_{i-1}}\left(\frac{1}{2}\frac{\epsilon_{i}\cdot\epsilon_{i+1}}{k_{i}\cdot k_{i+1}}(k_{i}-k_{i+1})\cdot k_{i-1}+\frac{\epsilon_{i}\cdot k_{i+1}}{k_{i}\cdot k_{i+1}}\epsilon_{i+1}\cdot k_{i-1}\right),\notag
\end{align}
valid in the polarization gauge
\begin{align}
\label{iepsilongauge}
\epsilon_i\cdot k_{i-1}=0~~~,~~~\epsilon_{i+1}\cdot k_{i+2}=0.
\end{align}
Despite its first glance appearance, the double-soft factor (\ref{iCHY2gluon}) is not manifestly anti-symmetric under $i+2 \leftrightarrow i-1$, since this symmetry is broken by the gauge choice (\ref{iepsilongauge}). This is consistent with the results of the previous section.\\
Therefore, the particular computation above for legs $1$ and $2$ going soft in an $n$-point amplitude gives the following factorization in the double soft gluon limit
\begin{align}
\label{CHY2gluon}
\lim_{p_1 \sim p_2 \to 0} \mathcal{A}_{n}\to& S^{(0)}_{1,2}\mathcal{A}_{n-2}\\
=&\left[\frac{1}{(k_{1}+k_{2})\cdot k_{3}}\left(\frac{1}{2}\frac{\epsilon_{1}\cdot\epsilon_{2}}{k_{1}\cdot k_{2}}(k_{1}-k_{2})\cdot k_{3}-\frac{\epsilon_{2}\cdot k_{1}}{k_{1}\cdot k_{2}}\epsilon_{1}\cdot k_{3}\right)\right.\notag\\
&\left.-\frac{1}{(k_{1}+k_{2})\cdot k_{n}}\left(\frac{1}{2}\frac{\epsilon_{1}\cdot\epsilon_{2}}{k_{1}\cdot k_{2}}(k_{1}-k_{2})\cdot k_{n}+\frac{\epsilon_{1}\cdot k_{2}}{k_{1}\cdot k_{2}}\epsilon_{2}\cdot k_{n}\right)\right]\mathcal{A}_{n-2},\notag
\end{align}
valid in the gauge (\ref{epsilongauge}). Here we emphasize again that since the above result is obtained from the CHY formula, it features only physical poles, it holds in arbitrary dimension and for all helicity combinations of the two soft gluons.\\
\indent Let us now compare (\ref{CHY2gluon}) to the result (\ref{eq:tnjw}) obtained from BCFW. Specifying to four dimensions and selecting $(1^+,2^-)$ helicities for the soft gluons, we use the following standard dictionary to translate $R^3_{1,2}$ and $R^n_{1,2}$ into spinor helicity formalism:
\begin{align}
k_i\cdot k_j = \frac{1}{2}\langle ij\rangle [ji]~~,~~\epsilon^{+}_1\cdot k_i =\frac{[1i]\langle in\rangle}{\sqrt{2}\langle n1\rangle}~~,~~\epsilon^{-}_2\cdot k_i =\frac{\langle 2i\rangle[i3]}{\sqrt{2}[23]}~~,~~\epsilon^{+}_1\cdot\epsilon^{-}_2 =\frac{\langle 2n\rangle[13]}{[23]\langle n1\rangle}.
\end{align}
Here we have selected proper reference spinors to account for the gauge (\ref{epsilongauge}).\footnote{Note that the specific choice of reference spinors merely facilitates the proper conversion of the result (\ref{CHY2gluon}) to spinor helicity formalism. Once the conversion is done, full gauge invariance is recovered for the final result in spinor helicity language i.e. (\ref{eq:tnjw2}).} Anticipating that $R^3_{1,2}$ and $R^n_{1,2}$ roughly correspond to the two terms that are summed in (\ref{eq:tnjw}), we notice that $R^3_{1,2}$ already features an $s_{123}\approx 2 k_3\cdot(k_1+k_2)$ and $R^n_{1,2}$ an $s_{n12}\approx 2 k_n\cdot(k_1+k_2)$ in the denominator. So in both cases we introduce an extra factor of $\langle n| 1+2 |3]$ in numerator and denominator, and expand the numerators. The Schouten identity then yields a slight simplification such that the terms in the numerators separate into an expected part and a part proportional to $s_{123}$ or $s_{n12}$ in the two cases respectively. Finally, subtracting the resulting $R^n_{1,2}$ from $R^3_{1,2}$ displays some cancellation and we are left with exactly the terms appearing in (\ref{eq:tnjw}).\footnote{One should keep in mind that in spinor helicity formalism factors of $\sqrt{2}$ from the amplitude are absorbed into the coupling constant in front. In case of the double soft gluon factor this amounts to an overall extra factor of $2$ which is suppressed in (\ref{eq:tnjw}).} 

Similarly, we can show that selecting the soft gluons to be of the same helicity, i.e. $(1^+,2^+)$, the double soft gluon factor (\ref{CHY2gluon}) reduces to the product of two single soft factors. Here we also use:
\begin{align}
\epsilon^{+}_2\cdot k_i =\frac{[2i]\langle i3\rangle}{\sqrt{2}\langle 32\rangle}~~,~~\epsilon^{+}_1\cdot\epsilon^{+}_2 =\frac{\langle n3\rangle[21]}{\langle n1\rangle\langle 32\rangle}.
\end{align}
In this case no strategic term manipulations are needed. $R^3_{1,2}$ directly reduces to half of the expected result and $R^n_{1,2}$ to minus half of it, so that $(R^3_{1,2}-R^n_{1,2})$ properly gives what we expect.  

\section{Triple-soft gluons} \label{section:triplegluons}

With results of the double-soft limit at hand, we can go on to study the universal behavior of scattering amplitudes with multiple gluons being soft. Here we will take a look at the triple-soft limit, which is a natural next step beyond the double-soft limit. Again the non-trivial cases occur when all soft gluons are adjacent. Beside the straightforward case of all soft gluons having the same helicity, there are two helicity configurations of interest: $A(1^+, 2^-, 3^-, \ldots)$ and $A(1^+, 2^-, 3^+, \ldots)$, where $1,2$ and $3$ are the soft legs. 

Let us begin with the first case, $A(1^+, 2^-, 3^-, \ldots)$. It is easy to see that the following BCFW diagrams are dominant in the soft limit 
\beq
\vcenter{\hbox{\includegraphics[scale=0.6]{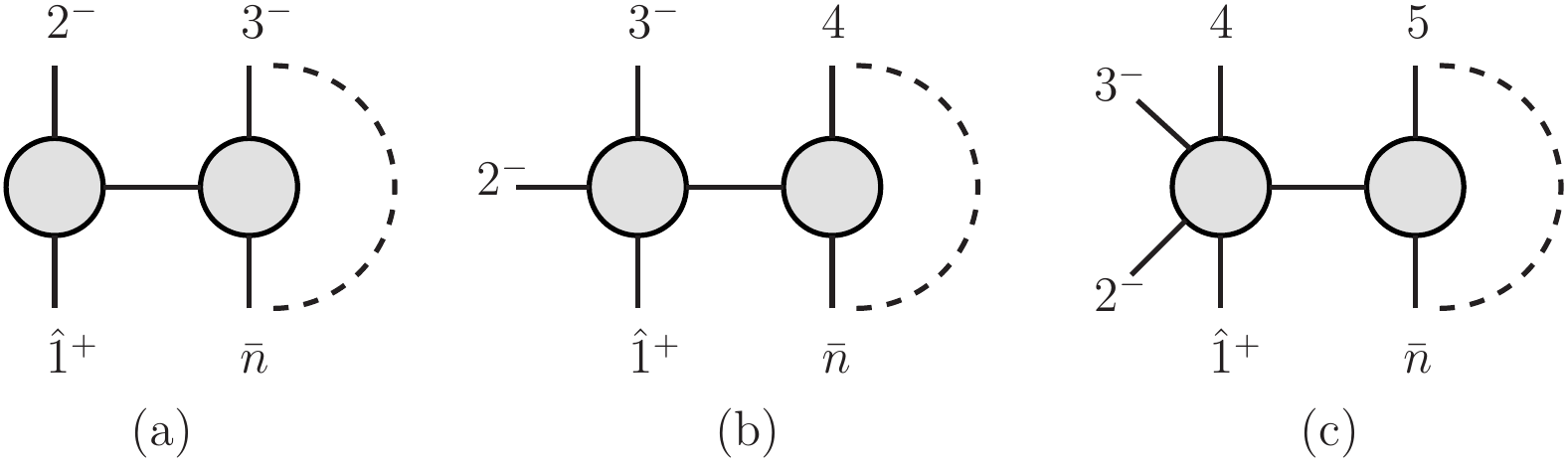}}}
\eeq
Since the calculation is similar to that of the double-soft limit, we will be brief here. The contribution from diagram (a) to the soft factor gives 
\beqa
\mathcal{S}^{+--}_{\rm (a)} = {[ \hat{P}1]^3 \over[12] [ 2 \hat{P}] s_{12}  } { [ \bar{n} 3] \over [\bar{n} \hat{P}][ \hat{P} 3]}{ [ \bar{n} 4] \over [ \bar{n} 3][34]} \, ,
\eeqa
where we have used the fact that $\hat{P}$ is soft, as well as the result of the double-soft limit with two negative-helicity gluons. Specifying $\hat{P}$ in terms of external momenta, the soft factor simplifies to
\beqa
\mathcal{S}^{+--}_{\rm (a)} = { \langle n2\rangle^3 [n4] \over \langle n1 \rangle \langle 12 \rangle [34] \langle n|K_{12}|3] s_{n12} }\,,
\eeqa
where $K_{i \ldots j} = k_i + \ldots + k_j$. Similarly, diagram (b) gives
\beqa
\mathcal{S}^{+--}_{\rm (b)} = {[ \hat{P} 1]^3 \over [12][23][3 \hat{P}] s_{123} } { [  \bar{n} 4 ] \over [ \bar{n} \hat{P} ][ \hat{P} 4 ] }=  
{[4n] \langle n| K_{23}|1]^3   \over [12][23] \langle n|K_{12} |3] \langle n| K_{123} |4] s_{123}s_{n123} }\, .
\eeqa
Finally, for diagram (c) a couple of remarks are in order.  First we note that if gluon $4$ has positive helicity, there are two allowed cases for the helicity of the internal line in the BCFW diagram. However, it is clear that the one diagram with an NMHV five-point amplitude on the left-hand side is dominant in the soft limit.  Second, just as in the case of the double-soft limit, the result is independent of the helicity of gluon $4$. Therefore, we can choose it to be negative and conclude
\beqa
\mathcal{S}^{+--}_{\rm (c)} = {  [\hat{P}1 ]^3 \over [12][23][34][4 \hat{P} ] s_{1234}} = 
{ [14]^3 \langle n4 \rangle \over [12][23][34] \langle n|K_{123}|4] s_{1234}} \,.
\eeqa
Summing over the three contributions, we obtain the universal behavior of amplitudes with three adjacent soft gluons 
\beqa
A(1^+, 2^-, 3^-, 4, \ldots,n)\big{|}_{p_1 \sim p_2 \sim p_3 \rightarrow 0} \rightarrow \left( \mathcal{S}^{+--}_{\rm (a)}+ \mathcal{S}^{+--}_{\rm (b)} 
+ \mathcal{S}^{+--}_{\rm (c)} \right) A_{n-3} \,.
\eeqa
Now we go on to consider the second case of interest, $A(1^+, 2^-, 3^+, \ldots)$. The result is given by the same set of BCFW diagrams, but now with the helicity of gluon $3$ changed 
\beq
\vcenter{\hbox{\includegraphics[scale=0.6]{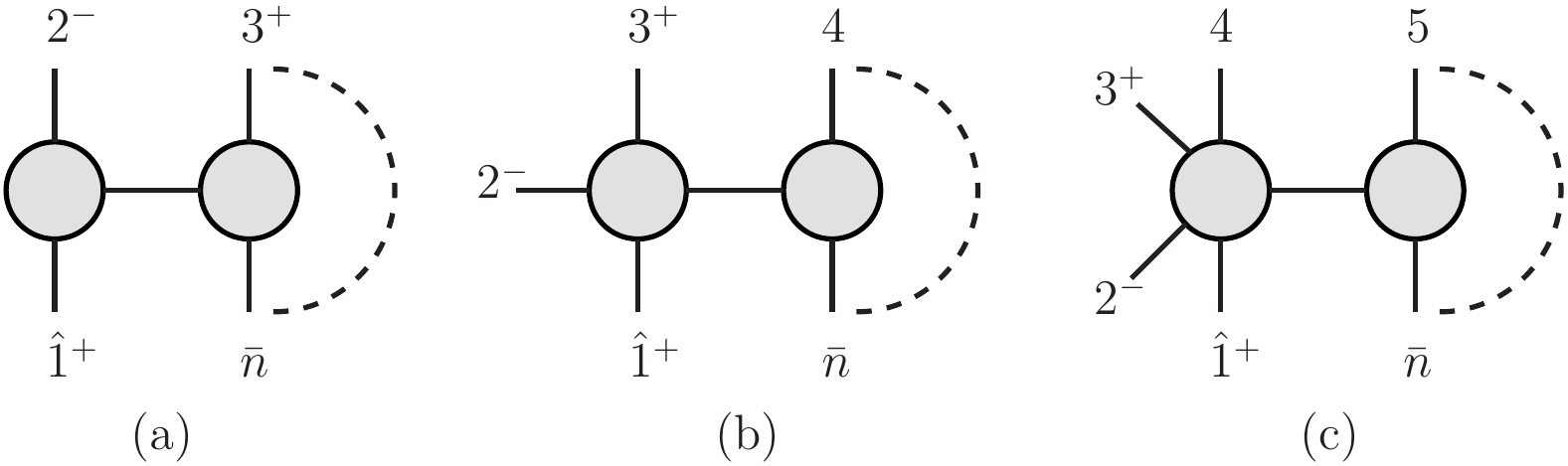}}}
\eeq
From diagram (a) we have
\beqa
\mathcal{S}^{+-+}_{\rm (a)} = { [ \hat{P}1 ]^3 \over [12][ 2 \hat{P}] s_{12}  } {1 \over \langle 4| \hat{P} + 3| \bar{n}] } 
\left( { \langle \hat{P} 4 \rangle^3 [ \bar{n} 4] \over \langle \hat{P} 3 \rangle \langle 34 \rangle s_{\hat{P} 34} } +
{ [ \bar{n} 3 ]^3 \langle {n} 4 \rangle \over [ \bar{n} \hat{P} ] [ \hat{P}  3 ] s_{\bar{n} \hat{P} 3} }  \right)\,,
\eeqa
where we have applied the alternating helicity double-soft gluon theorem (\ref{eq:tnjw}) to the right sub-amplitude in the BCFW diagram. After some further simplifications taking the soft limit into account, we obtain 
\beqa 
\mathcal{S}^{+-+}_{\rm (a)} &=& { \langle 2n\rangle^3 \over \langle n1\rangle \langle 12\rangle 
\langle 2|K_{n1}K_{n123}|4 \rangle } \cr
&\times&
\left( { \langle 24\rangle^3   [n4] \over \langle 23 \rangle \langle 34\rangle \langle 2|K_{34}K_{1234}|n \rangle } + { \langle n2\rangle \langle n4\rangle [n3]^3 \over \langle n|K_{12}|3] s_{n12}s_{n123} }  \right)\,. 
\eeqa
Note that it is not allowed to discard the soft momenta $k_{1}$, $k_{2}$ and $k_{3}$ in $\langle 2|K_{n1}K_{n123}|4 \rangle$ and $\langle 2|K_{34}K_{1234}|4 \rangle$ to further simplify the above expressions in the soft limit. For the diagram (b) we find
\beqa
\mathcal{S}^{+-+}_{\rm (b)} = {[13]^4 \over [12][23][3 \hat{P}] [ \hat{P} 1] s_{123} } { \langle n4 \rangle \over \langle n \hat{P} \rangle \langle \hat{P} 4 \rangle }=  
{[13]^4 \langle n4 \rangle  \over [12] [23] \langle n|K_{12} |3] \langle 4|K_{23} |1] s_{123} }\,.
\eeqa
Finally, diagram (c) gives
\beqa
\mathcal{S}^{+-+}_{\rm (c)}  = {  \langle 24\rangle^4 \over \langle \hat{1} 2\rangle\langle 23\rangle \langle 34\rangle \langle 4 \hat{P}\rangle \langle \hat{P} \hat{1} \rangle s_{1234} } = 
{ \langle 24\rangle^4  [41]^3 \langle n 4 \rangle \over 
\langle 23\rangle \langle 34\rangle  \langle 4|K_{23}|1] \langle n|K_{1234}K_{34}|2 \rangle s_{234} s_{1234}} \,. 
\eeqa
In conclusion, we obtain the following soft theorem for three adjacent soft gluons with alternating helicities
\beqa
A(1^+, 2^-, 3^+, 4, \ldots,n)\big{|}_{p_1 \sim p_2 \sim p_3 \rightarrow 0} \rightarrow  \left( \mathcal{S}^{+-+}_{\rm (a)}+ \mathcal{S}^{+-+}_{\rm (b)} 
+ \mathcal{S}^{+-+}_{\rm (c)} \right) A_{n-3} \,.
\eeqa
Before we close this section, we would like to remark that both soft factors $\sum_{i={\rm (a),(b),(c)}}\mathcal{S}^{+--}_i$ and $\sum_{i={\rm (a),(b),(c)}}\mathcal{S}_i^{+-+}$ nicely reduce to a product of a single-soft factor and a double-soft factor if we take any one of the three soft gluons to be soft first. Finally, we note that all the unphysical poles appear in pairs, and we have checked numerically that they all precisely cancel at leading order in the soft limit. 

\section{Multi-soft gravitons}\label{section:gravity}

In this section we comment that, unlike in the case of two soft gluons, the double-soft-graviton limit is simply given by the product of two single-soft gravitons, independent of their helicity configuration. For instance, let us consider soft gravitons of opposite helicity $g^{++}_1$ and $g^{--}_2$. Similar to the case of double-soft gluons from BCFW recursion, one needs to consider the following three diagrams:
\beq
\vcenter{\hbox{\includegraphics[scale=0.6]{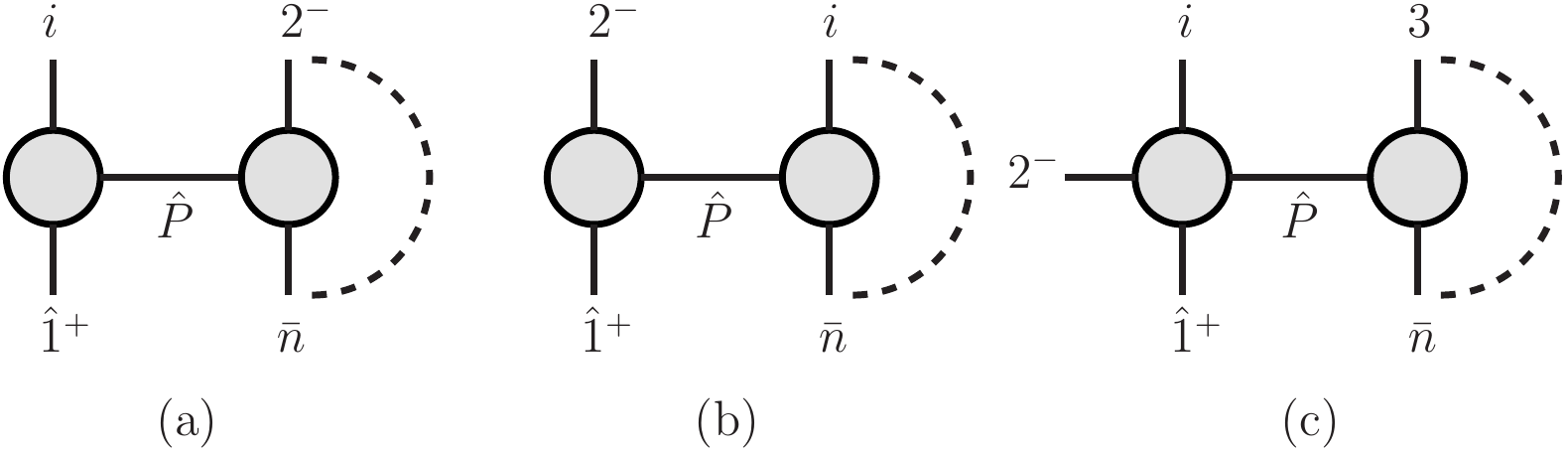}}}
\eeq
In fact, a simple analysis of three- and four-point amplitudes reveals that only the diagram (a) will contribute at leading order in the double-soft limit. A simple way to obtain the result for diagram (a) is to view it as an ``inverse-soft" diagram~\cite{inversesoft}, where leg $1^+$ is considered as being added to an $(n{-}1)$-point amplitude making use of
\beqa \label{diagrama}
M^{\rm (a)}_n 
&=& \sum_{i \neq 2} \mathcal{S}_{1^+}(i) M_{n-1}(i',  \ldots, 2^-, \ldots, n' ). \,
\eeqa
Here the soft factor $\mathcal{S}_{1^+}$ is defined as
\beq
\mathcal{S}_{1^+}(i) = { {  \langle n i\rangle^2  [i 1 ]\over \langle n 1\rangle^2  \langle i1\rangle  } } \, .
\eeq
In this diagram the shifted legs are $p_{i'}$ and $p_{n'}$, which are given by
\beqa
\tilde{ \lambda }_{i'} = \tilde{ \lambda }_{i} + { \langle 1n\rangle \over \langle in\rangle  } \tilde{ \lambda }_{1}\,,\quad
\tilde{ \lambda }_{n'} = \tilde{ \lambda }_{n} + { \langle 1i\rangle \over \langle ni\rangle  } \tilde{ \lambda }_{1} \, .
\eeqa
In the soft limit we simply have $p_{i'} \rightarrow p_i$ and $p_{n'} \rightarrow p_n$. Since $p_2$ is soft as well, it follows from the single-soft graviton theorem that the above expression reduces to
\beqa \label{diagramaa}
M_n 
& \rightarrow & \sum_{i \neq 2} \mathcal{S}_{1^+}(i) \sum_{j \neq 1} \mathcal{S}_{2^-}(j) M(3, \ldots, n) 
\eeqa
with 
\beq
\mathcal{S}_{2^-}(j) = { {  [ x j ] [ y j ] \langle j 2 \rangle \over [ x 2 ] [ y2 ] [ j2 ]  } } 
\eeq
for any choices of $x$ and $y$. Considering that $M^{\rm (a)}_n$ is the dominant diagram at leading order, we have replaced it by the full tree-level amplitude $M_n$. Finally, we note that the terms $\mathcal{S}_{1^+}(2)$ and $\mathcal{S}_{2^-}(1)$, which are missing in the summation in (\ref{diagramaa}), are subleading in the limit. Thus the result can be alternatively written as
\beqa \label{diagramaaa}
M_n  
& \rightarrow & \sum_{i} \mathcal{S}_{1^+}(i) \sum_{j \neq 1} \mathcal{S}_{2^-}(j) M(3, \ldots, n) \cr
& \sim & \sum_{j} \mathcal{S}_{2^-}(j) \sum_{i \neq 2} \mathcal{S}_{1^+}(i)  M(3, \ldots, n)\cr
& \sim & \sum_{i,\, j} \mathcal{S}_{1^+}(i) \mathcal{S}_{2^-}(j)  M(3, \ldots, n) \,,
\eeqa
being simply the product of two single-soft factors. As mentioned earlier, this confirms that the leading double-soft-graviton limit can be obtained by taking the gravitons to be soft in succession, in either order, unlike the case of double-soft gluons. Given the result of double-soft gravitons, it is straightforward to see that it can be extended to the case of multiple soft gravitons, such that the soft factor of multiple-soft gravitons should be given by the product of multiple single-soft-graviton factors for any number of soft gravitons.

\section{Double-soft limits in supersymmetric gauge theories} \label{section:SYM}

In this section, we move on to study the universal behavior of scattering amplitudes in supersymmetric gauge theories (in particular $\mathcal{N}=4$ SYM and pure $\mathcal{N}=2$ SYM) in the limit with the momenta of two scalars or two fermions being soft. The double-soft-scalar limit was first studied in $\mathcal{N}=8$ supergravity in~\cite{nima}, where the $70$ scalar fields in the theory parametrize the coset space $E_{7(7)}/{\rm SU(8)}$. Thus these scalar fields behave as ``pions". As pointed out in~\cite{nima}, amplitudes in this theory vanish in the single-soft-scalar limit consistent with the famous ``Adler's zero"~\cite{adler}, 
and behave universally in the double-soft-scalar limit in a manner analogous to the soft-pion theorem 
\beq \label{nimasimplest}
\lim_{\tau \to0} M_n\left( \phi^{I I_1 I_2 I_3}(  \tau p_1),\; \phi_{J I_1 I_2 I_3}( \tau p_2),3,\cdots,n \right)
 \rightarrow
{1 \over 2}\sum_{i=3}^{n} \frac{p_i\cdot (p_1-p_2)}{p_i\cdot(p_1+p_2)}(R_{i})^I\,_J M_{n-2} \,,
\eeq
where $(R_{i})^I\,_J$ is the generator for SU$(8)$ rotations on particle $i$
\beq
(R_{i})^I\,_J = \eta^I_i { \partial_{\eta^J_i}  } \, .
\eeq
Recently, this result was extended to more general supersymmetric gravity theories~\cite{congkao2}, including $4 \leq \mathcal{N} <8$ supergravity theories in four dimensions as well as $\mathcal{N} =16$ supergravity in three dimensions. 
Soft-scalar theorems have been very useful in determining the UV counter terms in supergravity theories~\cite{closeE77, congkao2}.
It is known that for supersymmetric gauge theories (in particular $\mathcal{N}=4$ SYM), a generic vacuum has mostly massive particles, and the massless S-matrix only exists at the origin of moduli space. Thus one should not expect that the scalars would behave as ``pions". Indeed it is easy to see that the amplitudes in $\mathcal{N}=4$ SYM do not vanish in the single-soft-scalar limit, in contrast to supergravity theories. However, as in~\cite{congkao1}, one can still ask whether the amplitudes in SYM exhibit some universal behavior in certain soft limits. This is what we will explore in this section. 

\subsection{Double-soft scalars in $\mathcal{N}=4$ SYM}

The on-shell fields in $\mathcal{N}=4$ SYM can be nicely packaged into a superfield \cite{Nair:1988bq},
\beq
\mathcal{A}(\eta) = g^+ + \eta^A \psi_A + {1 \over 2!} \eta^A \eta^B \phi_{AB} + 
{1 \over 3!} \eta^A \eta^B \eta^C \psi_{ABC} + (\eta^1\eta^2 \eta^3 \eta^4) g^- \, ,
\eeq
where $g^+$ is the positive-helicity gluon, $\psi_A$ is the spin $+1/2$ gluino, and so on. In this section, we will consider the limit with two scalars $\phi_{AB}$ becoming soft. First of all, as we mentioned previously, it is easy to see that amplitudes in $\mathcal{N}=4$ SYM behave as $\mathcal{O}(\tau^0)$ in the single-soft-scalar limit. 

Let us now consider the double-soft-scalar limit. First we note that when the two soft scalars are not adjacent, the amplitude is not singular, and thus it cannot behave universally under the soft limit. So we will only consider the case where the two soft scalars are adjacent, which is singular and therefore universal. To be precise, we take $p_1$ and $p_2$ to be soft. Furthermore, if two scalars have no common SU$(4)$ index, they form a singlet and the leading singular result should simply be given by the single-soft gluon limit. However, as pointed out in~\cite{congkao2} for supergravity theories, one can extract interesting information about this case by introducing suitably anti-symmetrised amplitudes. This is particularly relevant to pure $\mathcal{N}=2$ SYM where two scalars can only form a singlet, which will be discussed in the next section. Here we will focus on the case where two scalars do not form a singlet, as was considered in~\cite{nima} for $\mathcal{N}=8$ supergravity. For this configuration it is easy to see that the leading contribution arises when two soft scalars have one and only one common SU$(4)$ index. In terms of the BCFW representation of the amplitude, there are two leading contributions in the double-soft limit:
\beq \label{BCFWN4soft}
\vcenter{\hbox{\includegraphics[scale=0.6]{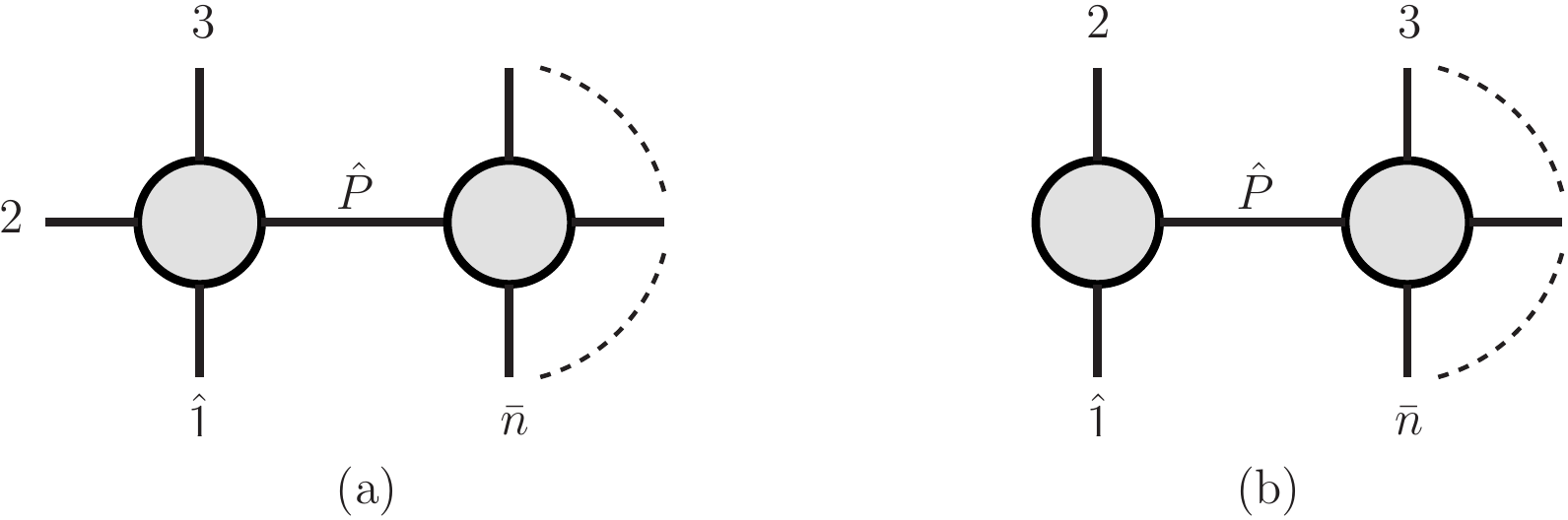}}}
\eeq
After integrating out $\eta_{\hat{P}}$, the contribution (a) is given by
\beqa
A_{\rm (a)}&=& 
\int d^4 \eta_1 d^4\eta_2  \eta^{A}_1 \eta^B_1 \eta^{B}_2 \eta^C_2
{ \langle \hat{1} \hat{P} \rangle^4 
\delta^{(4)}( \eta_1 + {\langle \hat{P} 2 \rangle \over \langle \hat{P} \hat{1} \rangle } \eta_2 
+ {\langle \hat{P} 3 \rangle \over \langle \hat{P} \hat{1} \rangle} \eta_3 )  
\over \langle \hat{1} 2\rangle \langle 23\rangle \langle 3 \hat{P} \rangle \langle \hat{P} \hat{1} \rangle  s_{123}}  
{\rm exp} \left( - { \langle \hat{1} 2\rangle \over \langle \hat{1} \hat{P}\rangle }\eta_2 { \partial \over \partial \eta_3} \right) \cr
&\times &
{\rm exp} \left(  z_P \eta_1 { \partial \over \partial \eta_n } \right)  A_{n-2} 
\eeqa
where the integration over $\eta$'s selects the soft legs $1$ and $2$ to be scalars. Note that, as mentioned above, we are interested in the case where two scalars have one common SU$(4)$ index. We have applied the super BCFW recursion relations~\cite{nima, Brandhuber:2008pf}, with shifts chosen as
\beq
\lambda_{\hat{1}} = \lambda_1 - z \lambda_n \, , \quad 
\tilde{\lambda}_{\bar{n}} =\tilde{\lambda}_n + z \tilde{\lambda}_1\, , \quad 
\eta_{\bar{n} } =  \eta_n + z \eta_1 \, .
\eeq
Finally, we have written the shifts in $A_{n-2}$ in an exponentiated form and only kept the leading terms. There are two possible ways to get a leading contribution above, one is by expanding $\eta_2$ from ${\rm exp} \left( - { \langle \hat{1} 2\rangle \over \langle \hat{1} \hat{P}\rangle }\eta_2 { \partial \over \partial \eta_3} \right)$, and another one is by expanding $\eta_1$ from ${\rm exp} \left(  z_P \eta_1 { \partial \over \partial \eta_n } \right)$. In the first case we get one $\eta_2$ from the exponent, thus from the fermionic delta-function $\delta^{(4)}$ we have one $\eta_2$, two $\eta_1$'s and one $\eta_3$. Thus we obtain, 
\beqa
A_{\rm (a),1} = 
{ \langle \hat{1} \hat{P} \rangle^4 
\over \langle \hat{1} 2\rangle \langle 23\rangle \langle 3 \hat{P} \rangle \langle \hat{P} \hat{1} \rangle  s_{123}}  
{ \langle \hat{1} 2\rangle \over \langle \hat{1} \hat{P}\rangle } {\langle \hat{P} 2 \rangle \over \langle \hat{P} \hat{1} \rangle }
{\langle \hat{P} 3 \rangle \over \langle \hat{P} \hat{1} \rangle} 
\eta^B_3 { \partial \over \partial \eta^D_3}   A_{n-2}\,,
\eeqa
where an extra minus due to the fermionic integral has been included. In the soft limit the above expression simplifies to
\beqa
A_{\rm (a),1} \rightarrow
{ 1
\over  2 p_3 \cdot (p_1 + p_2) }   
\eta^B_3 { \partial \over \partial \eta^D_3}   A_{n-2}\,.
\eeqa
Analogously, we obtain the second contribution, which is given by
\beqa
A_{\rm (a),2}&=& 
{ \langle \hat{1} \hat{P} \rangle^4 
\over \langle \hat{1} 2\rangle \langle 23\rangle \langle 3 \hat{P} \rangle \langle \hat{P} \hat{1} \rangle  s_{123}}  
z_P  \left( {\langle \hat{P} 2 \rangle \over \langle \hat{P} \hat{1} \rangle } \right)^2
{\langle \hat{P} 3 \rangle \over \langle \hat{P} \hat{1} \rangle} 
\eta^B_3 { \partial \over \partial \eta^D_n}   A_{n-2}\cr
& \rightarrow & 
{ 1 
\over   \langle n | 1+2 |3 ] }    
\eta^B_3 { \partial \over \partial \eta^D_n}   A_{n-2}\,,
\eeqa
where we used the on-shell solution $z_P = -{ s_{123} \over \langle n | 3+2 |1 ] } \sim -{ s_{123} \over \langle n 3 \rangle [31] }$. 

Let us now consider the diagram (b), for which a similar consideration leads to
\beqa \label{figb}
A_{\rm (b)} = { \delta^{(4)} \left( [12]\eta_{P} + [2 \hat{P}] \eta_1 + [\hat{P} 1] \eta_2 \right) \over [12][2 \hat{P}] [ \hat{P} 1 ]s_{12} }
{\rm exp} \left(  z_P \eta_1 { \partial \over \partial{\eta_n} } \right)  A_{n-1}(\hat{P}, \ldots, n) \,. 
\eeqa
Now, using the fact that $\hat{P}$ is also soft, one can apply the supersymmetric single-soft theorem to $A_{n-1}(\hat{P}, \ldots, n)$. Thus we have
\beqa
A_{n-1}(\hat{P}, 3, \ldots, n)\big{|}_{\hat{P} \rightarrow 0 } \rightarrow {[n3] \over [\hat{P} 3][n \hat{P}] } 
\delta^{(4)} \left( \eta_P + {[n \hat{P}] \over [3n]} \eta_3 +{[\hat{P} 3] \over [3n]}  \eta_{n} \right) A_{n-2} \, . 
\eeqa
Substituting this result into eq.~(\ref{figb}), integrating out $\eta_P$, and selecting the scalar components we find
\beqa 
A_{\rm (b) }&=&- { [2 \hat{P}][\hat{P} 1]^2  z_P \over [12][2 \hat{P}] [ \hat{P} 1 ]s_{12} }
{[n3] \over [\hat{P} 3][n \hat{P}] } 
\left({[12] [n \hat{P}] \over [3n]} \eta^B_3 +{[12] [\hat{P} 3] \over [3n]}  \eta^B_{n} \right) { \partial \over \partial{\eta^D_n} } A_{n-2}
\cr
& \rightarrow & 
-\left( {1\over \langle n|1+2|3]} \eta^B_3 +{1 \over 2 p_n \cdot (p_1 + p_2) }  \eta^B_{n} \right) { \partial \over \partial{\eta^D_n} } A_{n-2}\,.
\eeqa
We observe that the unphysical pole cancels out. In particular, the first term in $A_{\rm (b) }$ cancels $A_{\rm (a),2 }$, and we obtain the double soft-scalar theorem in $\mathcal{N}=4$ SYM 
\beqa \label{doublescalars}
A_n( (\phi_1)_{CD}, (\phi_2)^{BC}, \ldots ) \rightarrow \left({ 1 \over 2p_3 \cdot (p_1 + p_2)} \eta^B_3 \partial_{\eta^D_3}  -
{ 1 \over 2p_n \cdot (p_1 + p_2)} \eta^B_n \partial_{\eta^D_n} \right) A_{n-2} \, ,
\eeqa
where $ \phi^{BC} = \epsilon^{ABCD} \phi_{DA}$. Note the appearance of the R-symmetry generators
$\eta^B \partial_{\eta^D}$. As mentioned earlier, although scalars in SYM are not Goldstone bosons, we find that our result very much resembles what has been found in $\mathcal{N}=8$ supergravity. 
Furthermore, as we will see, the double-soft-scalar theorem is exact even when we consider amplitudes in open superstring theory, meaning that it does not receive any $\alpha'$ corrections from string theory. Finally, we remark that the subleading order of this limit will be finite and thus not universal, since general BCFW diagrams start to contribute. This is the same for the double-soft limit of scalars in $\mathcal{N}=8$ supergravity. 


\subsection{Double-soft scalars in pure $\mathcal{N}=2$ SYM}

In this section we consider the double-soft-scalar limit for pure $\mathcal{N}=2$ SYM. Due to the fact that it is not a maximally supersymmetric theory, the on-shell fields in $\mathcal{N}=2$ SYM are separated into two distinct mulitplets. These multiplets can be nicely obtained from $\mathcal{N}=4$ SYM by SUSY truncation~\cite{Elvang:2011fx},
\beq
\mathcal{A}^{\mathcal{N}=2}(\eta) = \mathcal{A}^{\mathcal{N}=4}(\eta)\big{|}_{\eta^3, \eta^4 \rightarrow 0 } \, , \quad
\bar{\mathcal{A}}^{\mathcal{N}=2}(\eta) =\int d \eta^3 d \eta^4 \mathcal{A}^{\mathcal{N}=4}(\eta) \,, 
\eeq
where $\mathcal{A}^{\mathcal{N}=4}(\eta)$ is the superfield in $\mathcal{N}=4$ SYM that we defined in the previous section. Therefore, we see that the scalar in $\mathcal{A}^{\mathcal{N}=2}(\eta)$ corresponds to $\phi_{12}$ in $\mathcal{N}=4$ SYM, while the scalar in $\bar{\mathcal{A}}^{\mathcal{N}=2}(\eta)$ corresponds to $\phi_{34}$ in $\mathcal{N}=4$ SYM. Thus they form a singlet. 

Since the scattering amplitudes in pure $\mathcal{N}=2$ SYM can be obtained from  amplitudes in $\mathcal{N}=4$ SYM via SUSY reduction, we will use the same strategy as in~\cite{congkao2}: instead of studying the amplitudes in $\mathcal{N}=2$ SYM directly we will study the relevant amplitude in $\mathcal{N}=4$ SYM first, and then reduce it to $\mathcal{N}=2$ SYM via the SUSY reduction. Now, in contrast with the case we studied in the previous section, here we are interested in precisely the amplitudes with the two soft scalars forming a singlet $A((\phi_1)_{12},(\phi_2)_{34}, \ldots )$, and with the following anti-symmetrization as introduced in~\cite{congkao2}:
\beq \label{anti-symmetrization}
A((\phi_1)_{12},(\phi_2)_{34}, \ldots ) - A((\phi_1)_{34},(\phi_2)_{12}, \ldots ) \, .
\eeq
Let us focus on $A((\phi_1)_{12},(\phi_2)_{34}, \ldots )$.  As before, in the soft limit  the dominant contributions are given by the diagrams shown in Fig.(\ref{BCFWN4soft}). The diagram (a) is given by a similar expression to that used above, but now we select different species of scalars  
\beqa
A_{\rm (a)}&=& 
\int d^4 \eta_1 d^4\eta_2  \eta^{1}_1 \eta^2_1 \eta^{3}_2 \eta^4_2
{ \langle \hat{1} \hat{P} \rangle^4 
\delta^{(4)}( \eta_1 + {\langle \hat{P} 2 \rangle \over \langle \hat{P} \hat{1} \rangle } \eta_2 
+ {\langle \hat{P} 3 \rangle \over \langle \hat{P} \hat{1} \rangle} \eta_3 )  
\over \langle \hat{1} 2\rangle \langle 23\rangle \langle 3 \hat{P} \rangle \langle \hat{P} \hat{1} \rangle  s_{123}}  
{\rm exp} \left( - { \langle \hat{1} 2\rangle \over \langle \hat{1} \hat{P}\rangle }\eta_2 { \partial \over \partial \eta_3} \right) \cr
&\times &
{\rm exp} \left(  z_P \eta_1 { \partial \over \partial \eta_n } \right)  A_{n-2}(\hat{3}, \ldots, \hat{n}) \, ,
\eeqa
here we keep BCFW shifted legs (shifting the momenta $p_3$ and $p_n$) in $A_{n-2}$, since we select different scalars, the leading term now comes from taking two $\eta_1$'s as well as two $\eta_2$'s from the fermionic delta-function, these shifted legs contribute in the subleading orders. However, all these contributions vanish after the anti-symmetrization (\ref{anti-symmetrization}). In fact, all the terms with all $\eta_1$'s and $\eta_2$'s from the fermionic delta-function vanish after the anti-symmetrization. Thus we will focus on terms with one $\eta_1$ or $\eta_2$ from the exponent, and the calculation proceeds as outlined in the previous section. Therefore, we  only quote the results
\beqa
A_{(a)} = \sum^2_{A=1} \left( {1 \over 2p_3 \cdot (p_1 + p_2)} \eta^A_3 { \partial \over \partial_{ \eta^A_3 }} +
{1 \over \langle n|1+2|3] }  \eta^A_3 { \partial \over \partial_{ \eta^A_n }} \right) A_{n-2} \,.
\eeqa
Similarly from diagram (b), we find
\beqa 
A_{\rm (b) }=
-\sum^2_{A=1}\left( {1\over \langle n|1+2|3]} \eta^A_3 +{1 \over 2 p_n \cdot (p_1 + p_2) }  \eta^A_{n} \right) { \partial \over \partial{\eta^A_n} } A_{n-2}\,.
\eeqa
Summing over all contributions, we find that after the anti-symmetrization we end up with
\beqa
 && A^{\mathcal{N}=4}_n( (\phi_1)_{12}, (\phi_2)_{34}, \ldots )- 
A^{\mathcal{N}=4}_n( (\phi_1)_{34}, (\phi_2)_{12}, \ldots )\big{|}_{p_1 \sim p_2 \rightarrow 0 } 
\cr
 &&= \left( \mathcal{S}^{\mathcal{N}=4}_{12;3} -\mathcal{S}^{\mathcal{N}=4}_{12;n }-\mathcal{S}^{\mathcal{N}=4}_{34;3} + \mathcal{S}^{\mathcal{N}=4}_{34;n} \right) A_{n-2} \,,
\eeqa
where the double-soft factor $\mathcal{S}^{\mathcal{N}=4}_{ij;k}$ is defined as
\beq
\mathcal{S}^{\mathcal{N}=4}_{ij;k} = \sum_{A=i,j} { 1 \over 2p_k \cdot (p_1 + p_2)} \eta^A_k \partial_{\eta^A_k} \, .
\eeq
Now we have to project this to $\mathcal{N}=2$ SUSY. The soft factor $\mathcal{S}^{\mathcal{N}=4}_{12;k}$ is unchanged, while $\mathcal{S}^{\mathcal{N}=4}_{34;k}$ depends on whether particles $3$ and $n$ are in the ${\mathcal{A}}^{\mathcal{N}=2}(\eta)$ or the $\bar{\mathcal{A}}^{\mathcal{N}=2}(\eta)$ multiplet. If they are in ${\mathcal{A}}^{\mathcal{N}=2}(\eta)$, then the contribution from $\mathcal{S}^{\mathcal{N}=4}_{34;k}$ should be discarded, since we set $\eta^3$ and $\eta^4$ to $0$. If they are in ${\mathcal{A}}^{\mathcal{N}=2}(\eta)$, then integrating out $\eta^3$ and $\eta^4$ the contribution from $\mathcal{S}^{\mathcal{N}=4}_{34;k}$ simply reduces to $2$. The result can be summarized as
\beqa
 A^{\mathcal{N}=2}_n( \phi_1, \bar{\phi}_2, \ldots )- 
A^{\mathcal{N}=2}_n( \bar{\phi}_1, \phi_2, \ldots )\big{|}_{p_1 \sim p_2 \rightarrow 0 } 
= \left(  R^{\mathcal{N}=2}_3 - R^{\mathcal{N}=2}_n \right)  A_{n-2} \, ,
\eeqa
where the U(1) generator $R^{\mathcal{N}=2}_i$ is defined as
\beq
R^{\mathcal{N}=2}_i = \sum^2_{I=1} \eta^I_i {\partial \over \partial_{\eta^I_i}}-2 \,, \quad ({\rm for} \,\,\, i \in {\mathcal{A}}^{\mathcal{N}=2} )\,, \quad 
R^{\mathcal{N}=2}_i = \sum^2_{I=1} \eta^I_i {\partial \over \partial_{\eta^I_i}} \,, \quad ({\rm for} \,\,\, i \in \bar{{\mathcal{A}}}^{\mathcal{N}=2} )\,,
\eeq
which precisely correspond to the U(1) part of the R-symmetry generators in pure $ \mathcal{N}=2$ SYM. 

\subsection{Double-soft fermions in $ \mathcal{N}=4$ and pure $ \mathcal{N}=2$ SYM} 

In a similar fashion one can study the limit with two soft fermions in $\mathcal{N}=4$ SYM as well as pure $\mathcal{N}=2$ SYM. As before, the interesting case occurs when the two fermions are adjacent. Because the (anti)-symmetrization procedure does not work for the double-soft fermions since they have different helicities~\cite{congkao1}, we will only consider the case when two fermions do not form a singlet. Thus the leading singular terms arise from adjacent fermions having one and only one common SU$(4)$ index. To be precise we take soft particles as $(\psi_1)_{D}$ and $(\psi_2)_{BCD}$. The calculation in terms of BCFW recursion relations is very similar to the case of double-soft scalars, and again the relevant BCFW diagrams are shown in Fig.(\ref{BCFWN4soft}). Let us quote them here for convenience
\beq
\vcenter{\hbox{\includegraphics[scale=0.6]{N4soft}}}
\eeq
As before, any other generic BCFW diagrams are subleading, since they are diagrams with a single-soft fermion and behave as $1/\sqrt{\tau}$ in our soft limit. In contrast, the dominant diagrams above behave as $1/{\tau}$. The contribution from diagram (a) is given by
\beqa
A_{\rm (a)}&=& 
\int d^4 \eta_1 d^4\eta_2   \eta^A_1 \eta^{B}_1 \eta^C_1 \eta^{A}_2
{ \langle \hat{1} \hat{P} \rangle^4 
\delta^{(4)}( \eta_1 + {\langle \hat{P} 2 \rangle \over \langle \hat{P} \hat{1} \rangle } \eta_2 
+ {\langle \hat{P} 3 \rangle \over \langle \hat{P} \hat{1} \rangle} \eta_3 )  
\over \langle \hat{1} 2\rangle \langle 23\rangle \langle 3 \hat{P} \rangle \langle \hat{P} \hat{1} \rangle  s_{123}}  
{\rm exp} \left(-  { \langle \hat{1} 2\rangle \over \langle \hat{1} \hat{P}\rangle }\eta_2 { \partial \over \partial \eta_3} \right) \cr
&\times &
{\rm exp} \left(  z_P \eta_1 { \partial \over \partial \eta_n } \right)
   A_{n-2} \, .
\eeqa
Now the integration on $\eta$'s is such that the soft legs $1$ and $2$ are the soft fermions of interest. Following the analysis of double-soft scalars, we find two kinds of contributions from diagram (a). One of them is given by
\beqa
A_{\rm (a),1}&=& 
{ \langle \hat{1} \hat{P} \rangle^4 
\over \langle \hat{1} 2\rangle \langle 23\rangle \langle 3 \hat{P} \rangle \langle \hat{P} \hat{1} \rangle  s_{123}}  
{ \langle \hat{1} 2\rangle \over \langle \hat{1} \hat{P}\rangle } 
{\langle \hat{P} 3 \rangle \over \langle \hat{P} \hat{1} \rangle} 
\left({\langle \hat{P} 2 \rangle \over \langle \hat{P} \hat{1} \rangle } \right)^2
\eta^A_3 { \partial \over \partial \eta^D_3}   A_{n-2}\cr
&=& 
-{ 1
\over  2 p_3 \cdot (p_1 + p_2) }  { [3 1] \over [3 2] } 
\eta^A_3 { \partial \over \partial \eta^D_3}   A_{n-2} \, ,
\eeqa
and the other contribution is
\beqa
A_{\rm (a),2}&=& 
{ \langle \hat{1} \hat{P} \rangle^4 
\over \langle \hat{1} 2\rangle \langle 23\rangle \langle 3 \hat{P} \rangle \langle \hat{P} \hat{1} \rangle  s_{123}}  
z_P  \left( {\langle \hat{P} 2 \rangle \over \langle \hat{P} \hat{1} \rangle } \right)^3
{\langle \hat{P} 3 \rangle \over \langle \hat{P} \hat{1} \rangle} 
\eta^A_3 { \partial \over \partial \eta^D_n}   A_{n-2}\cr
&=& 
-{ 1 
\over   \langle n | 1+2 |3 ] } { [3 1] \over [3 2] }   
\eta^A_3 { \partial \over \partial \eta^D_n}   A_{n-2} \, .
\eeqa
Similarly, from diagram (b) we find
\beqa 
A_{\rm (b) }&=& -{  [\hat{P} 1]^3  z_P \over [12][2 \hat{P}] [ \hat{P} 1 ]s_{12} }
{[n3] \over [\hat{P} 3][n \hat{P}] } 
\left({[12] [n \hat{P}] \over [3n]} \eta^A_3 +{[12] [\hat{P} 3] \over [3n]}  \eta^A_{n} \right) { \partial \over \partial{\eta^D_n} } A_{n-2}\cr
& \rightarrow & 
-{ \langle n 2 \rangle \over \langle n 1 \rangle } \left( {1\over \langle n|1+2|3]} \eta^A_3 + {1 \over 2 p_n \cdot (p_1 + p_2) }  \eta^A_{n} \right) { \partial \over \partial{\eta^D_n} } A_{n-2} \, .
\eeqa
Adding the results of the two diagrams together, we finally obtain the double soft-fermion theorem in $\mathcal{N}=4$ SYM, 
\beqa \label{doublefermions} \nonumber
 A_n( (\psi_1)_D, (\psi_2)^{A}, \ldots ) 
\rightarrow -{1\over [23]\langle n 1 \rangle } \left( { \langle n2\rangle  [23] \over 2p_n \cdot (p_1 + p_2)}  \eta^A_n \partial_{\eta^D_n} +{ \langle n 1 \rangle [13] \over 2p_3 \cdot (p_1 + p_2)}  \eta^A_3 \partial_{\eta^D_3}
+  \eta^A_3 { \partial \over \partial{\eta^D_n} }  \right) A_{n-2} \, ,
\\
\eeqa
Unlike the case of double-soft scalars, the cross term $\eta^A_3 { \partial \over \partial{\eta^D_n} }$ does not cancel anymore. However, all the unphysical poles cancel out manifestly. Note that the fermions in $\mathcal{N} =2$ SYM are not required to form a singlet like the scalars. The extension to the fermions in $\mathcal{N} =2$ SYM is straightforward via SUSY truncation as we discussed in the previous section.

\section{Double-soft limit in open superstring theory} \label{section:string}

It is known that the soft-scalar theorems in $\mathcal{N}=8$ supergravity are violated in the closed superstring theory if $\alpha'$ corrections are included~\cite{closeE77}. It is then natural to ask whether the newly established double-soft-scalar theorems in SYM would receive any $\alpha'$ corrections for scattering amplitudes in open superstring theory. We find remarkably that amplitudes in open superstring theory satisfy exactly the same double-soft-scalar theorems as in SYM theory. 

A general $n$-point color-ordered open string superamplitude of SYM vector multiplet at tree level can be very nicely expressed in terms of a basis of $(n{-}3)!$ SYM amplitudes~\cite{Mafra:2011nw, Mafra:2011nv}, 
\bea \label{generalopen}
\mathcal{A}(1,2, \ldots, n) = 
\sum_{\sigma \in S_{n-3}} F^{(2_{\sigma }, \ldots, (n{-}2)_{\sigma } )} 
A_{\rm SYM} (1, 2_{\sigma }, \ldots, (n{-}2)_{\sigma }, n{-}1, n)
\eea
where $A_{\rm SYM} (1, 2_{\sigma }, \ldots, (n{-}2)_{\sigma }, n{-}1, n)$ is the color-ordered tree-level amplitude of SYM, and the multiple hypergeometric functions are given as
\bea \nonumber
F^{( 2, \ldots, n{-}2)} = (-1)^{n-3} \int^1_{0< z_i < z_{i+1}} \prod^{n-2}_{j=2} dz_j \left( \prod_{i<l}  |z_{i l}|^{s_{il}} \right)
\left(  \prod^{[n/2]}_{k=2} \sum^{k-1}_{m=1} {s_{mk}  \over z_{mk}} \right)
 \left( \prod^{n-2}_{k=[n/2]+1} \sum^{n-1}_{m=k+1} {s_{km}  \over z_{km}}   \right) \, .
 \\
\eea
The Mandelstam variables are defined as $s_{ij}\equiv \alpha' (k_i{+}k_j)^2$. Here we have fixed the SL$(2, \mathbb{C})$ symmetry by choosing $z_1=0, z_{n-1}=1$ and $z_n = \infty$. Explicit expressions for the multiple hypergeometric functions in terms of $\alpha'$ expansion may be found in~\cite{Mafra:2011nv, Mafra:2011nw}. For instance, at four points we have, 
\bea
F^{(2)} &=& - \int^1_{0} dz_2 \, z^{s_{12}}_2 (1-z_2)^{s_{23}} \, {s_{12} \over z_{12} } = 
{\Gamma(1+s_{12}) \Gamma(1 + s_{23}) \over  \Gamma(1 +s_{12} + s_{23})}\cr
&=&
1- \zeta_2 s_{12} s_{23} + \zeta_3 s_{12} s_{13}s_{23} + \cdots \, .
\eea
Let us start with the six-point amplitude as a simple example, the string amplitude is given as
\bea 
\mathcal{A}(1,2, \ldots, 6) = 
\sum_{\sigma \in S_{3}} F^{(2_{\sigma }, 3_{\sigma }, 4_{\sigma } )} 
A_{\rm SYM} (1, 2_{\sigma }, 3_{\sigma }, 4_{\sigma }, 5, 6) \, .
\eea
It turns out to be convenient to take the soft limit on legs $3$ and $4$, more generally for a $n$-point amplitude, we take $p_{n-3}$ and $p_{n-2}$ to be soft. From the definition of $F^{( 2, \ldots, n{-}2)}$ (for $n=6$, the explicit expressions for $F^{(2_{\sigma }, 3_{\sigma }, 4_{\sigma } )}$ in $\alpha'$ expansion can be found in eq.(2.29) in~\cite{Mafra:2011nw}), it is easy to see that for six points only $F^{(2,3,4)}$ contributes in the limit, and it simply becomes $F^{(2)}$. Thus we have
\beqa 
\mathcal{A}(1,2, \ldots, 6) \rightarrow  
F^{(2)} 
\mathcal{S}_{ij} A_{\rm SYM} (1, 2, 5, 6)=
\mathcal{S}_{ij} \mathcal{A}(1,2,5, 6)
\, ,
\eeqa
where for the convenience of following discussion we defined the soft factor $\mathcal{S}_{ij}$
\beq
\mathcal{S}_{ij} = { 1 \over 2p_i \cdot (p + q)} \eta^I_i \partial_{\eta^J_i} -
{ 1 \over 2p_j \cdot (p + q)} \eta^I_j \partial_{\eta^J_j} \, ,
\eeq
with $p$ and $q$ being the soft legs. In the above case these are $p_3$ and $p_4$. The amplitude with a general multiplicity can be considered similarly by following a proof of single-soft-gluon theorem in~\cite{Mafra:2011nw}. First of all, we note that only those permutations $\sigma \in S_{n-3}$ where indices $(n{-}3)$ and $(n{-}2)$ are adjacent may contribute, since otherwise the amplitudes in SYM would be finite and therefore subleading. By the property of hypergeometric functions $F$, the position $(n{-}4)$ should always be on the left of $(n{-}3)$ and $(n{-}2)$. Furthermore, $(n{-}3)$ and $(n{-}2)$ should be in the canonical order, meaning that $(n{-}3)$ should be on the left of $(n{-}2)$. Otherwise, for all the above cases the multiple hypergeometric function $F$ is vanishing. For such $\sigma$'s we find the following configurations: 
\begin{itemize}
\item $\sigma \in S_{n-5}$ with $(n-4)_{\sigma} = n-4$, we have
\beqa \label{situation1}
&& F_{n}^{(\sigma)}A_{\rm SYM}(1, 2_{\sigma}, \ldots, (n-4), (n-3), (n-2), (n-1), n ) \cr
&&\rightarrow 
\mathcal{S}_{n-4, n-1} F_{n-2}^{(\sigma)} A_{\rm SYM}(1, 2_{\sigma}, \ldots, (n-4),  (n-1), n )
\eeqa 
In the following, we then consider the cases with $(n-4)_{\sigma} \neq n-4$. 
\item $\sigma \in S_{n-5}$ with $(n-4)_{\sigma} \neq n-4$, we have
\beqa \label{situation2}
&& F_{n}^{(\sigma)} A_{\rm SYM}(1, 2_{\sigma}, \ldots, (n-4)_{\sigma}, (n-3), (n-2), (n-1), n ) \cr
&&\rightarrow 
\mathcal{S}_{(n-4)_{\sigma}, n-1} F_{n-2}^{(\sigma)} A_{\rm SYM}(1, 2_{\sigma}, \ldots, (n-4)_{\sigma},  (n-1), n )
\eeqa
\item Finally, we have the non-vanishing contribution with $\sigma \in S_{n-5}$ with $(n-4)_{\sigma} \in \{ 2_{\sigma}, \ldots, i_{\sigma} \} $,
\beqa \label{situation3}
&& F_{n}^{(\sigma)} A_{\rm SYM}(1, 2_{\sigma}, \ldots, i_{\sigma}, (n-3), (n-2), (i+1)_{\sigma}, \ldots, (n-4)_{\sigma}, (n-1), n ) \cr
&&\rightarrow 
\mathcal{S}_{i_{\sigma}, (i+1)_{\sigma}} F_{n-2}^{(\sigma)} A_{\rm SYM}(1, 2_{\sigma}, \ldots, (n-4)_{\sigma},  (n-1), n )
\eeqa
\end{itemize}
Using the definition of the soft factor $\mathcal{S}_{ij}$ (in particular its antisymmetric property), we find that the results of the second the third cases combine nicely, 
\beqa
{\rm eq.(\ref{situation2})} + {\rm eq.(\ref{situation3})} = \mathcal{S}_{n-4, n-1} F_{n-2}^{\sigma} A_{\rm SYM}(1, 2_{\sigma}, \ldots, (n-4)_{\sigma},  (n-1), n ) \, .
\eeqa
Combining with the result of (\ref{situation1}), this concludes the proof that the amplitudes in open superstring theory satisfy the same double-soft-scalar theorem as in SYM theory. 

\section*{Acknowledgement}
We would like to thank Steven Avery, Massimo Bianchi, Burkhard Schwab, Marcus Spradlin and Gabriele Travaglini for useful conversations. 
AV and MZ are supported by the US Department of Energy
under contract DE-FG02-11ER41742 Early Career Award and the Sloan Research Foundation.


\end{document}